\documentclass[twocolumn,showpacs,preprintnumbers,superscriptaddress,amsmath,amssymb,nofootinbib]{revtex4}

\input psfig.sty

\usepackage{epsfig}

\usepackage{graphicx}
\usepackage{dcolumn}
\usepackage{bm}

\newcommand{\be}{\begin{equation}}
\newcommand{\ee}{\end{equation}}

\begin{document}

\title{Decaying Vacuum Cosmology and its Scalar Field Description}

\author{F. E. M. Costa} \email{ernandesmc@usp.br}

\author{J. A. S. Lima} \email{limajas@astro.iag.usp.br}

\author{F. A. Oliveira} \email{foliveira@astro.iag.usp.br}

\affiliation{Departamento de Astronomia, Universidade de S\~ao Paulo, Rua do Mat\~ao 1226,
05508-900, S\~ao Paulo, SP, Brazil}

\date{\today}

\begin{abstract}

We discuss the cosmological consequences of an interacting model in the dark sector in which the $\Lambda$ component evolves as a truncated power series of the Hubble parameter.
In order to constrain the free parameters of the model we carry out a joint statistical analysis involving observational data from current type Ia supernovae, recent estimates of the cosmic microwave background shift parameter and baryon acoustic oscillations measurements. Finally, we adopt  a theoretical method to derive the coupled scalar field version for this non-equilibrium decaying vacuum accelerating cosmology.

\end{abstract}

\maketitle

\section{Introduction}

In the last decade, the analysis and interpretation of observational data are strongly indicating that the present Universe is accelerating and that
its geometry is spatially flat \cite{Riess,Kowalski10,Spergel,Komatsu}.  These
observations gave rise to a resurgence of interest in a non-zero cosmological constant $\Lambda$ that dominates the current composition of the Universe \cite{rev1}.
The main motivation behind this idea is that a very small value for $\Lambda$ ($\simeq 70\%$ of the cosmic composition), not only would explain the observed late-time
acceleration of the Universe (as indicated by type Ia
supernovae observations  but would also reconcile the inflationary flatness prediction, corroborated by cosmic microwave background data, ($\Omega_{\rm{Total}}
\simeq 1$) with the current clustering estimates that point systematically to $\Omega_m \simeq 0.3$ \cite{Spergel,omega}.

Although flat models with a very small cosmological term are in good agreement with almost all sets of observational data, the value of ${\Lambda}$ inferred from observations
($\rho_{\Lambda} = \Lambda/8\pi G \sim 10^{-47} {\rm{GeV}}^4$) is between 50-120 orders of magnitude below estimates given by quantum field theory ($\rho_{\Lambda}
\sim 10^{71}$ ${\rm{GeV}}^4$). This extreme discrepancy originates an extreme fine-tuning problem, the so-called cosmological constant problem, and requires a complete
cancellation from an unknown physical mechanism \cite{Zee85,weinb,Sahni2000,CLW92,LM1}.

Another problem with the cosmological term is that, although a very small (but non-zero) value for $\Lambda$ could conceivably be explained by some unknown physical
symmetry being broken by a small amount, one should be able to explain not only why it is so small but also why it is exactly the right value that is just beginning
to dominate the energy density of the Universe now. Since both components (dark matter and cosmological term) are usually assumed to be independent and, therefore,
scale in different ways, this would require an unbelievable co-occurrence, usually referred to as the coincidence problem (CP).

Many phenomenological models with variable cosmological term (decaying vacuum) have been proposed in literature as an attempt to alleviate the cosmological constant problem
\cite{ozer,CLW92,LM1} and more recently the coincidence problem \cite{wang, Jesus, costa77, costa81}. In the context of the general relativity theory a cosmological
term that varies in space or time requires a coupling with some other cosmic component, so that the total energy-momentum tensor is conserved. Thus,
$\Lambda (t)$ models provide either a process of a particle production or a time-varying  mass of the dark matter particles increases \cite{alc05}.

In this paper, we focus our attention on the theoretical and observational aspects of a $\Lambda(t)$CDM model in which the cosmological term evolves as a truncated power series of
the Hubble parameter. The decaying vacuum component is assumed to be coupled only with the dark matter (interacting dark sector) thereby producing dark matter particles.  The work is structured as follows. In Section II, we discuss the general aspects and solve the basic equations of the decaying vacuum model. The
observational constraints on parameters of the model are discussed in Section III. In section IV, by following an approach proposed by Maia and Lima \cite{maia}, we also derive a coupled scalar field version for this vacuum decay scenario. Finally, the basic results are summarized  in the conclusion Section V.

\section{The $\Lambda(t)$ model}

Let us first consider a homogeneous, isotropic and spatially flat universe described by the Friedmann-Robertson-Walker line element. In such a background, the Einstein
field equations are given by
\begin{equation}\label{fri}
8\pi G \rho_f + \Lambda = 3H^2\;,
\end{equation}
and
\begin{equation}\label{fri2}
8\pi G p_f - \Lambda =-2\dot{H} - 3H^2\;,
\end{equation}
where $\rho_f$ and $p_f$ are, respectively, the energy density and pressure of the cosmic fluid (radiation, baryons and dark matter) and the dot denotes the derivative with respect to time.

In this paper, we will work with the $\Lambda(t)$ function written as a power series of the Hubble parameter up to the second order, i.e.,
\begin{equation}\label{evo}
\Lambda (H) =  \lambda + \sigma H + 3\beta H^2\;,
\end{equation}
where $\lambda$ and $\sigma$ are constants with dimensions of $H^{2}$ and $H$, respectively, while $\beta$ is a dimensionless constant. The term proportional to $H^{2}$ was proposed long ago by Carvalho, Lima and Waga \cite{CLW92} based on dimensional arguments. Later on, it was also justified
by using techniques from the renormalization group approach \cite{Sola}. The linear term was first discussed by Carneiro and collaborators \cite{saulo}. The interest in the extended model represented by the above truncated power series was also suggested in the appendix C of a paper by Basilakos, Plionis and Sol\'a \cite{BPS}. As remarked before, here we discuss the observational potentialities of this model, and, for completeness, we also derive the associated scalar field description.  

Now by combining the above equations and considering that $p_f = (\gamma - 1) \rho_f$ where $\gamma$ is the adiabatic index, we find
\begin{equation}
2\dot{H} + 3\gamma(1 -\beta)H^2 - \sigma \gamma H  - \gamma \lambda = 0\;.
\end{equation}
The integration of above equation leads the following $H$ parameter
\begin{equation}\label{hdet}
H \equiv \frac{\dot{a}}{a} = \frac{1}{3(1-\beta)} \left[\frac{\sigma}{2} + \frac{\alpha}{\gamma} \frac{e^{\alpha t} + 1}{e^{\alpha t} - 1}\right]\;,
\end{equation}
where $\alpha = (\gamma /2)\sqrt{\sigma^{2} -12\lambda (1-\beta)}$ and $a$ is the cosmic scale factor.
In what follows, we will restrict our analysis to case in which $\lambda = 0$. The practical reason for this choice is that the linear term in Eq. (\ref{evo}) causes a transition decelerating/accelerating phase. So that, we can obtain the following solution for the scale factor
\begin{equation}\label{scale}
a(t) = C (e^{\sigma \gamma t/2} - 1)^{2/3 \gamma (1- \beta)}\;,
\end{equation}
where $C$ is a integration constant. Combining Eqs. (\ref{hdet}) and (\ref{scale}) one finds an expression for $H(a)$, i.e.,
\begin{equation}\label{hdea}
H = \frac{\sigma}{3(1-\beta)} \left[1 + \left(\frac{C}{a}\right)^{3(1-\beta)/2} \right]\;,
\end{equation}
and we have used $\gamma = 1$ in the last equation. By taking Eqs. (\ref{fri}), (\ref{evo}) and (\ref{hdea}) at current time it is possible to show that $C = \Omega_{m}/(1- \beta - \Omega_{m})^{2/3(1-\beta)}$ and $\sigma = 3 H_0 (1-\beta - \Omega_{m})$, where $\Omega_{m} = 8\pi G\rho_{m,0} /3H_{0}^{2}$. Now, by considering that $a = (1+z)^{-1}$, the Hubble parameter can be rewritten as
\begin{equation}\label{friedmann}
{H}= H_0 \left[1 - \frac{\Omega_{m}}{{1-\beta}} + \frac{\Omega_{m}}{{1-\beta}}(1+z)^{3(1-\beta)/2}\right]\;.
\end{equation}
It should be noticed that for $\beta = 0$, the above expression reduces to the dynamical $\Lambda$ solution derived in Ref. \cite{saulo}.

It is also straightforward to show from (\ref{friedmann}) that the deceleration parameter, defined as $q (a) \equiv - a\ddot{a}/a^{2}$, now takes the following form:
\begin{equation}\label{qdez}
q(z) = \frac{3}{2} \frac{\Omega_{m}(1+z)^{3(1-\beta)/2}}{\left[1 - \frac{\Omega_{m}}{{1-\beta}} + \frac{\Omega_{m}}{{1-\beta}} (1+z)^{3(1-\beta)/2}\right]} - 1\;.
\end{equation}
Note that for $z = 0$, one finds  $q_0 = 1.5\Omega_{m} - 1$, and, therefore,  the present value of the deceleration parameter is negative (see Table 1) and does not depend on the values of $\beta$. It is easy to check that the transition redshift $(z_{t})$, defined to be the zero point of the deceleration parameter, is given by
\begin{equation}\label{zstar}
z_{t} =  \left[\frac{2(1 - \beta - \Omega_{m})}{(1-3\beta)\Omega_{m}}\right]^{2/3(1-\beta)} - 1\;.
\end{equation}

In this connection, we recall that some authors have recently suggested a picture based on theoretical and observational evidences where the cosmic
acceleration may have just peaked and started to slow down \cite{slow}. From expression  (\ref{qdez}) for the deceleration parameter, we see that in the future (negative redshifts) the value of q(z) approaches -1.

For the sake of completeness, we derive the age-redshift relation for this class of $\Lambda {\rm{(t)}}$ models. We find
\begin{equation}\label{age}
t(z) = \frac{2}{\sigma} \ln \left[1 + \frac{1- \beta -\Omega_{m}}{\Omega_{m}}(1 + z)^{-3(1-\beta)/2}\right]\;.
\end{equation}
Note also that the value of the present age of the Universe in this decaying vacuum model is given by
\begin{equation}
t_0 = \frac{2H_0^{-1}}{3(1- \beta -\Omega_{m})} \ln \left(\frac{1-\beta}{\Omega_{m}}\right)\;,
\end{equation}
so that $H_0t_0 \sim 1$ in agreement with the current observations (see Table 1).

\section{Observational analysis}

In this section we will discuss bounds on the free parameters $\Omega_{m}$ and $\beta$. To this end we will use different observational sets of data, as described below.

\subsection{Constraints from Type Ia supernovae}

In this test we use one of the most recent SNe Ia compilation, the so-called Union 2.1 sample compiled in Ref.~\cite{union21} which includes 580 data points after selection cuts.

The best fit to the set of parameters $s$ = ($\Omega_{m}$, $\beta$) is found by using a $\chi^2$ statistics, i.e.,
\begin{equation}\label{chisquare}
\chi^2 = \sum_{i=1}^N\frac{{[\mu_{p}^{i}(z|s)(z_i) - \mu_{o}^{i}(z|s)}]^{2}} {\sigma_i^2}\;,
\end{equation}
where $\mu_{p}^{i}(z|s) = 5\log d_L + 25$ is the predicted distance modulus for a supernova at redshift (z), $d_L$ is the luminosity distance, $\mu_{o}^{i}(z|s)$ is the extinction corrected distance modulus for a given SNe Ia at $z_i$ and $\sigma_i$ is the uncertainty in the individual distance moduli.

\subsection{Constraints from CMB/BAO ratio}

Additionally, we also use measurements derived from the product of the CMB acoustic scale $\ell_{A} = \pi d_A (z_*)/r_s(z_*)$ and from the ratio of the sound horizon scale at the drag epoch to the BAO dilation scale, $r_s(z_d)/D_V(z_{\rm{BAO}})$, where $d_A (z_*)$ is the comoving angular-diameter distance to recombination ($z_* = 1089$) and $r_s(z_*)$ is the comoving sound horizon at photon decoupling.
In the above expressions,  $z_d \simeq 1020$ is the redshift of the drag epoch (at which the acoustic oscillations are frozen in) and the dilation scale, $D_V$, is given by $D_V(z) = [zr^{2}(z)/H(z)]^{1/3}$. By combining the ratio
$r_s (z_d = 1020)/r_s (z_*=1090) = 1.044 \pm 0.019$ ~\cite{Komatsu,Percival} with the measurements of $r_s(z_d )/D_V(z_{\rm{BAO}})$ at $z_{\rm{BAO}} =$ 0.20, 0.35 and 0.6., one finds~\cite{sollerman,blake}
$$
f_{0.20} = d_A (z_*)/D_V (0.2) = 18.32 \pm0.59\;,
$$
$$
f_{0.35} = d_A (z_*)/D_V (0.35) = 10.55 \pm 0.35\;,
$$
$$
f_{0.6} = d_A (z_*)/D_V (0.60) = 6.65 \pm 0.32\;.
$$

\begin{figure}[t]
\label{plane}
\centerline{\psfig{figure=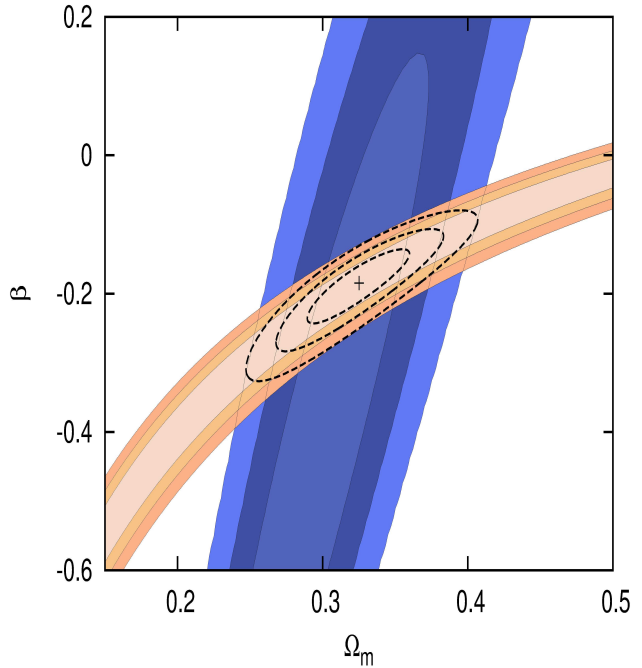,width=3.4truein,height=3.8truein,angle=0}}
\caption{The results of our statistical analysis. The constraints from SNe Ia and CMB/BAO ratio data are shown by shaded contours for 3$\sigma$ confidence level,
while the contours from combining observational data are represented with black lines for (68.3\% CL, 95.4\% and 99.73\%) confidence intervals.}
\end{figure}

\subsection{Combining observational sets data}

Let us now discuss the constraints given by the above different sets of data. In our statistical analysis,  we minimize the function $\chi^2_{\rm{T}} = \chi^2_{\rm{SN}} + \chi^2_{\rm{CMB/BAO}}$, where $\chi^2_{\rm{SN}}$ and $\chi^2_{\rm{CMB/BAO}}$ corresponding to the SNe Ia and CMB/BAO $\chi^2$ functions, respectively.

In Figure 1, we display the space parameter ($ \Omega_m, \beta$). The supernova data set fitted with Union 2.1 are shown by the bluish contours while the orange contours are derived from CMB/BAO ratio data alone. The results from our joint analysis are shown by dashed lines. By marginalizing on the nuisance parameter $h$ ($H_0=100hkm s^{-1}  Mpc^{-1}$) we find $\beta=-0.185^{+0.045+0.075+0.095}_{-0.055-0.085-0.135} $ and $\Omega_m = 0.325^{+0.035+0.060+0.080}_{-0.035-0.060-0.080}$ at $68.3\%$, $95.4\%$ and $99.7\%$ of confidence level, respectively, with $\chi^{2}_{min}=564$ and $\nu=581$ degrees of freedom. The reduced $\chi^{2}_{r}=0.97$ where ($\chi^{2}_{r}=\chi^{2}_{min} /\nu$), thereby showing that the model provides a very good fit to these data. We also note that, due to the process of matter production resulting from the vacuum decay, the current observational bounds on $\Omega_{m}$ provide a value slightly higher than the $\Lambda$CDM model.

By using the best-fit values of $\Omega_{m}$ and $\beta$ it  follows that $q_0 = -0.513{}^{+0.053}_{-0.053}$, $z_{t} =0.99{}^{+0.19}_{-0.20}$ and $t_0 H_0 = 1.003{}^{+0.049}_{-0.051}$. The main results of our joint analysis are shown in Table 1.

\begin{table}[t] \label{table:results}
\begin{center}  
\caption{The results from joint statistical analysis}
\begin{tabular}{ccccc}
\hline  \\
\multicolumn{1}{c}{Parameter} \quad &         \multicolumn{1}{c}{best fit}     \quad &        \multicolumn{1}{c}{error bars ${}^{+1\sigma}_{-1\sigma}$}& \vspace{0.2cm} \\ \hline \\
$\Omega_{m}$                                   	&  \quad$0.325$                      &        ${}^{+0.035}_{-0.035}$                       \vspace{0.2cm}\\
$\beta$                                       	&  $-0.185$                          &              ${}^{+0.045}_{-0.055}$                       \vspace{0.2cm}\\
$\sigma/H_0$                           		&  \quad$2.58$                       &            ${}^{+0.24}_{-0.27}$                    \vspace{0.2cm}\\
$q_0$                                		&  $-0.513$                          &          ${}^{+0.053}_{-0.053}$                    \vspace{0.2cm}\\
$z_{t}$                           		&  \quad$0.99$                       &          ${}^{+0.19}_{-0.20}$                    \vspace{0.2cm}\\
$t_0 H_0$                              		&  \quad$1.003 $                     &          ${}^{+0.049}_{-0.051}$                    \vspace{0.2cm}\\
\hline  \hline
\end{tabular}
\end{center}
\end{table}

\section{Scalar Field Description}

$\Lambda$(t) cosmologies constitute a possible way to address the cosmological constant problem. However, in the absence of a natural guidance from fundamental physics, one needs to specify a phenomenological time-dependence for $\Lambda$ in order to establish a definite model and study their observational and theoretical implications. A possible way to seek for physically motivated models is to represent them through a field theoretical language, the easiest way being through scalar fields. In this section,  we discuss a possible derivation of a time-varying $\Lambda$ models from fundamental physics based on the approach suggested in Ref. \cite{maia} (see also \cite{maia2} for a general description with arbitrary curvature).

Following standard lines \cite{maia},  let us define the function
\begin{equation}
\gamma_* \equiv - \frac{2\dot{H}}{3H^2} =  1 - \frac{\Lambda}{3H^2}\;.
\end{equation}
Each $\Lambda(t)$ model is specified through the $\gamma_{\star}$ time-dependent parameter. Thus, for the vacuum decay scenario
presented here, the function $\gamma_{\star}$ takes the following form:
\begin{equation}\label{gama}
\gamma_* = \frac{3(1-\beta)H^2 - \sigma H}{3H^2}\;.
\end{equation}

As an intermediate step, it is convenient to  replace the vacuum energy density and pressure as given by  Eqs. (\ref{fri}) and (\ref{fri2}) by the corresponding scalar field expressions, i.e.,
\begin{equation}\label{repla}
\Lambda/{8\pi G} \rightarrow \rho_{\phi}= \dot{\phi}^{2}/{2} +V(\phi)\;,
\end{equation}
and
\begin{equation}\label{repla2}
-\Lambda/{8\pi G} \to p_{\phi}= \dot{\phi}^{2}/{2} -V(\phi)\;,
\end{equation}
where $\rho_{\phi}$ and $p_{\phi}$ are, respectively, the energy density and pressure associated to the coupled scalar field $\phi$ whose potential is $V(\phi)$.

Now, by defining a parameter $x \equiv {\dot\phi^2}/{(\dot\phi^2 + \rho_{dm})}$ with $0 \leq x \leq 1$, we can manipulate Eqs. (\ref{fri}), (\ref{fri2}), (\ref{repla})
and (\ref{repla2}) to separate the scalar field contributions (see \cite{maia} for more details), i.e.,
\begin{equation}\label{eq:dotphi}
\dot{\phi}^2 = {3H^2 \over 8\pi G} \gamma_* x\;,
\end{equation}
and
\begin{equation}\label{vphi}
V(\phi) = {3H^2 \over 8\pi G} \left[1 - \gamma_* \left(1-{x\over
2}\right)\right]\;.
\end{equation}

Note that the above equations link directly the field and its potential with the related quantities of the dynamical $\Lambda{\rm{(t)}}$ case. From Eq. (\ref{eq:dotphi}),
one can show that  the field $\phi$ in terms of the Hubble parameter reads
\begin{eqnarray} \label{ph}
\phi - \phi_{0} = \pm \sqrt{\frac{3}{8\pi G}}\int_{H}^{H_{0}}{\sqrt{\frac{x}{\gamma_*}}\frac{dH}{H}}\;.
\end{eqnarray}

\begin{figure}[t]
\centerline{\psfig{figure=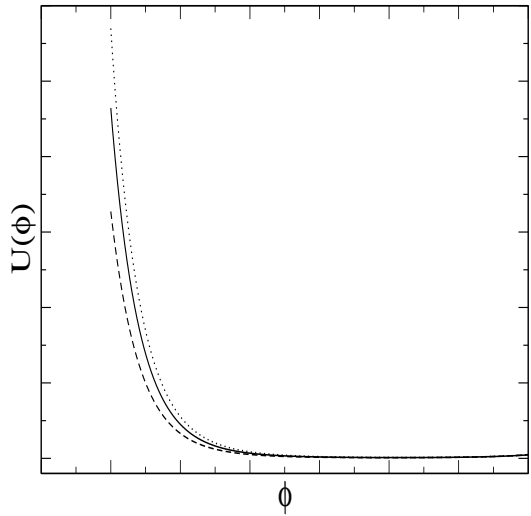,width=3.5truein,height=3.3truein,angle=-90}}
\caption{Behavior of the scalar field potential in units of the present critical energy density $U(\phi) \equiv V(\phi)/\rho_{c,0}$. The dotted, solid and dashed curves correspond to the lower, best fit and upper limits of the $\Omega_m$ and $\beta$ parameters at (1$\sigma$ C.L.). See discussion in the main text and Table 1.}
\end{figure}

Now, inserting Eq. (\ref{gama}) into Eq. (\ref{ph}) and integrating it, one obtains
\begin{equation}\label{fia2}
\phi - \phi_{0} = - \lambda^{-1} \ln\left[\frac{\sqrt{H - A} + \sqrt{H}}{\sqrt{H_{0} - A} + \sqrt{H_{0}}}\right]\;,
\end{equation}
where $\lambda = \sqrt{3 \pi G(1-\beta)/2x}$ and $A = \sigma/3(1-\beta)$. By using Eq. (\ref{vphi}) one finds
\begin{equation}\label{pot2}
V( \phi )  = \frac{D^{2}}{64\pi G}[(2-x)\sigma + Bf(\phi)]f(\phi)\;,
\end{equation}
where
\begin{equation}\label{pot3}
f(\phi) = \left[e^{-2\lambda (\phi - \phi_{0})} + \left(\frac{A}{D^{2}}\right)^{2}e^{2\lambda (\phi - \phi_{0})} + 2\frac{A}{D^{2}} \right],
\end{equation}
with $B = 3D^{2}[2\beta + (1-\beta)x]/4$ and $D = \sqrt{H_{0} - A} + \sqrt{H_{0}}$. Note that by assuming $\sigma = 0 \Longrightarrow A = 0$, and the potential takes the form of a simple exponential function thereby reducing to the potential initially found in Ref. \cite{maia}. It is worth mentioning that double exponential potentials of the type (\ref{pot2}) have been considered in the literature as viable examples of quintessence scenarios (see, e.g, \cite{dp}). As discussed in Ref.~\cite{dp1}, a scalar field potential given by the sum of two exponential terms is also motivated by dimensional reduction in M-theory with interesting implications for the late-time accelerating behavior of the cosmic expansion.

In Figure 2, we display the evolution of the scalar field potential (in units of the present critical energy density $[U( \phi )\equiv V(\phi)/\rho_{c,0}]$) for some selected values of the free parameters, namely: $\Omega_m = 0.29, 0.325, 0.36$ and $\beta = -0.24, -0.185, -0.14$. This choice correspond to the lower, best fit and upper limits on $\Omega_m$ and $\beta$ parameters at (1$\sigma$ C.L.). Note that all curves exhibit the same behavior differing only by a small shift.

\section{Conclusions}

In this paper we have investigated theoretical and observational features of a vacuum decay model in which the cosmological term is written as a power series of the Hubble
parameter up to the second order. We have performed a statistical analysis involving the lates observational measurements of SNe Ia, BAO peak and CMB shift
parameter and found stronger constraints on the parametric space $\Omega_{m} - \beta$. By using the best-fit values for $\Omega_{m} = 0.325{}^{+0.035}_{-0.035}$ and $\beta = -0.185{}^{+0.045}_{-0.055}$ we have also
derived the current values of $q_0$, $z_{t}$ and $t_0$, that are $q_0 =-0.513{}^{+0.053}_{-0.053}$, $z_{t} =0.99{}^{+0.19}_{-0.20}$ and $t_0 H_0 = 1.003{}^{+0.049}_{-0.051}$, respectively. These values are in excellent agreement with the predictions of the standard model (see Table 1).

By following a theoretical method developed in Ref. \cite{maia} we have also derived the scalar field description for this cosmology and shown that the present $\Lambda$(t)CDM model is identified with a coupled quintessence field with the potential of Eq. (\ref{pot2}). Since this $\Lambda$(t)CDM scenario is quite general (having many of the previous proposals as a particular case), we argue that a coupled quintessence field model whose potential is given by Eq. (\ref{pot2}) is dynamically equivalent to a large number of decaying vacuum scenarios previously discussed in the literature.

\begin{acknowledgments}

FEMC is supported by FAPESP under grants 2011/13018-0. JASL is partially supported by CNPq and FAPESP under grants 304792/2003-9 and 04/13668-0, respectively. FAO is supported by CNPq (Brazilian Research Agencies). The authors are grateful to V. C. Busti for helpful discussions.

\end{acknowledgments}

\end{document}